\documentclass[a4paper]{jpconf}
\usepackage{graphicx}
\usepackage{amsmath,amssymb}

\newcommand{\beq}{\begin{equation}}
\newcommand{\eeq}{\end{equation}}
\newcommand{\beqa}{\begin{eqnarray}}
\newcommand{\eeqa}{\end{eqnarray}}
\newcommand{\bseq}{\begin{subequations}}
\newcommand{\eseq}{\end{subequations}}

\def\x{{\boldsymbol x}}

\def\k{{\boldsymbol k}}

\def\p{{\boldsymbol p}}

\def\0{{\boldsymbol 0}}

\def\cal{\mathcal}
\def\lsim{\raise0.3ex\hbox{$<$\kern-0.75em\raise-1.1ex\hbox{$\sim$}}}
\def\gsim{\raise0.3ex\hbox{$>$\kern-0.75em\raise-1.1ex\hbox{$\sim$}}}

\begin{document}
\title{Heavy-flavor transport}

\author{Andrea Beraudo}

\address{INFN, Sezione di Torino, Via Pietro Giuria 1, I-10125 Torino}

\ead{beraudo@to.infn.it}

\begin{abstract}
The formation of a hot deconfined medium (Quark-Gluon Plasma) in high-energy nuclear collisions affects heavy-flavor observables. In the low/moderate-$p_T$ range transport calculations allow one to simulate the propagation of heavy quarks in the plasma and to evaluate the effect of the medium on the final hadronic spectra: results obtained with transport coefficients arising from different theoretical approaches can be compared to experimental data. Finally, a discussion of possible effects on heavy-flavor observables due to the possible formation of a hot-medium in small systems (like in p-A collisions) is presented.   
\end{abstract}

\section{Introduction}
Heavy quarks play an important and peculiar role in characterizing the medium produced in heavy-ion collisions.
The description of soft observables (low-$p_T$ hadrons) is based on hydrodynamics, assuming that one deals with a system (to which the measured particles belonged before decoupling) close to local thermal equilibrium.
Jet-quenching studies address the energy-degradation of external probes (high-$p_T$ partons) crossing the medium. The description of heavy-flavor observables on the other hand requires the employment/development of a setup (transport theory) allowing one to face a more general situation and in particular to describe how particles would (asymptotically) approach kinetic equilibrium with the medium in which they are embedded. At high $p_T$ the interest in heavy flavor is no longer related to their possible thermalization, but rather to the study of the mass and color-charge
dependence of jet-quenching: this last aspect however will not be addressed by our discussion. 

Why are charm and beauty quarks considered \emph{heavy}? This is due to the large value of their mass compared to other physical scales entering into the problem. First of all, $M\!\gg\!\Lambda_{\rm QCD}$, so that their initial hard production is under theoretical control, well described by pQCD. Secondly $M\gg T$, so that their thermal abundance is negligible and their final total multiplicity in the plasma (rapidly expanding and with a lifetime of at most $\sim$10 fm/c) created in the heavy-ion collisions is set by the initial hard production (possibly affected by nuclear corrections to the proton PDFs). Finally, in the context of a weakly-coupled relativistic plasma, their masses are also much larger than the typical momentum exchange with the medium particles, $M\gg gT$ ($g$ being the QCD coupling), such that many independent collisions are necessary to significantly change the trajectory/momentum of a heavy quark; this last observation will be of importance for the development of numerical schemes to simulate the propagation of heavy-flavor particles in the plasma.

The present contribution is organized as follows. In Sec.~\ref{sec:pp} we discuss how heavy-flavor production in elementary pp collisions can be described and nicely reproduced by pQCD. In Sec.~\ref{sec:AA} we address the case of nucleus-nucleus (A-A) collisions, where a hot Quark-Gluon Plasma (QGP) is expected to be produced and to modify heavy-flavor observables. Finally, in Sec.~\ref{sec:pA}, we move to small systems, like the ones produced in p-Pb or d-Au collisions, where, looking at soft observables, signatures of the formation of a medium displaying a collective behavior were found. Can this also affect heavy-flavor spectra? Some first estimates are provided.
\section{The benchmark: heavy flavor in pp collisions}\label{sec:pp}
\begin{figure}[ht]
\begin{center}
\includegraphics[clip,width=0.48\textwidth]{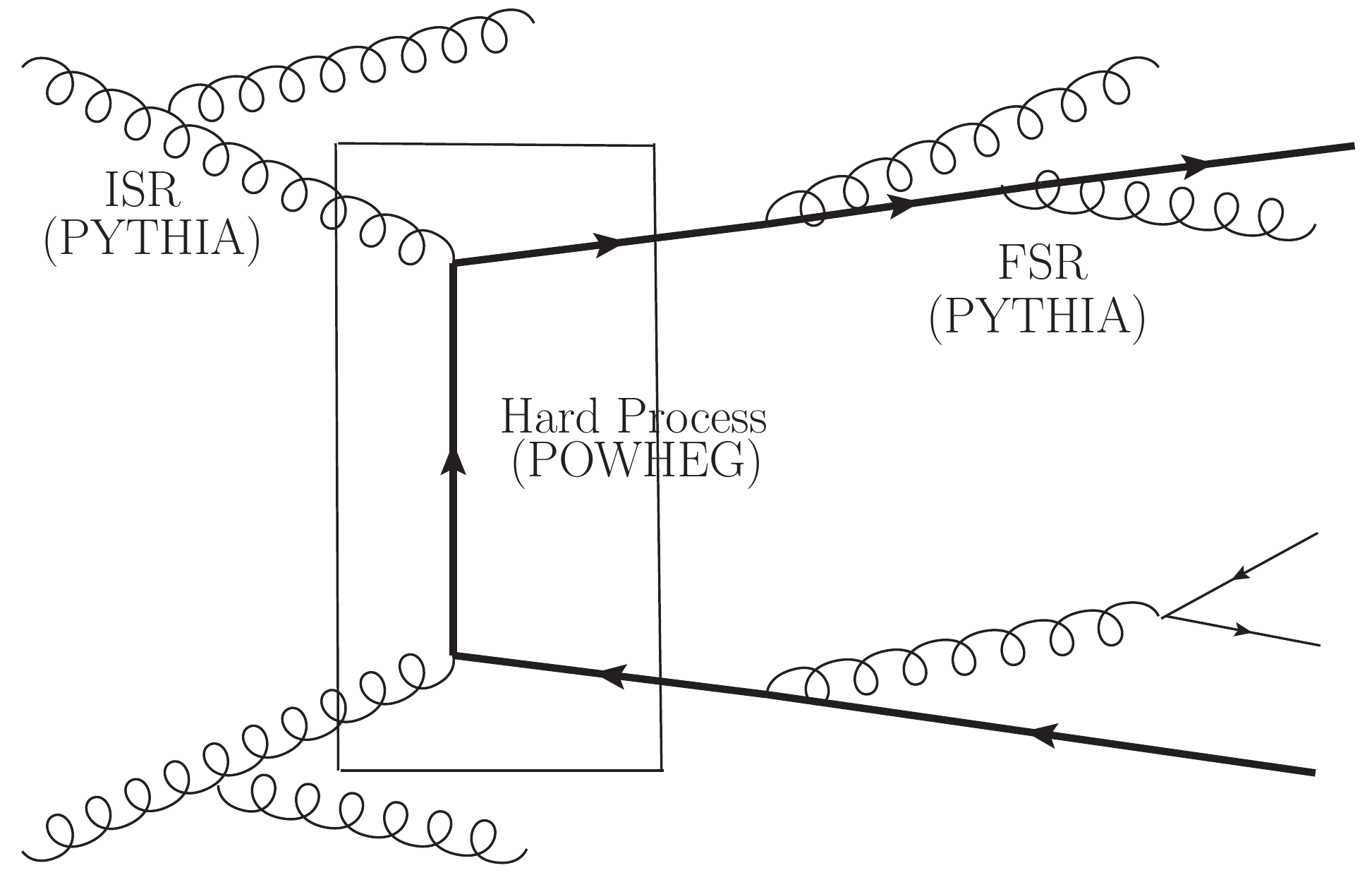}
\includegraphics[clip,width=0.48\textwidth]{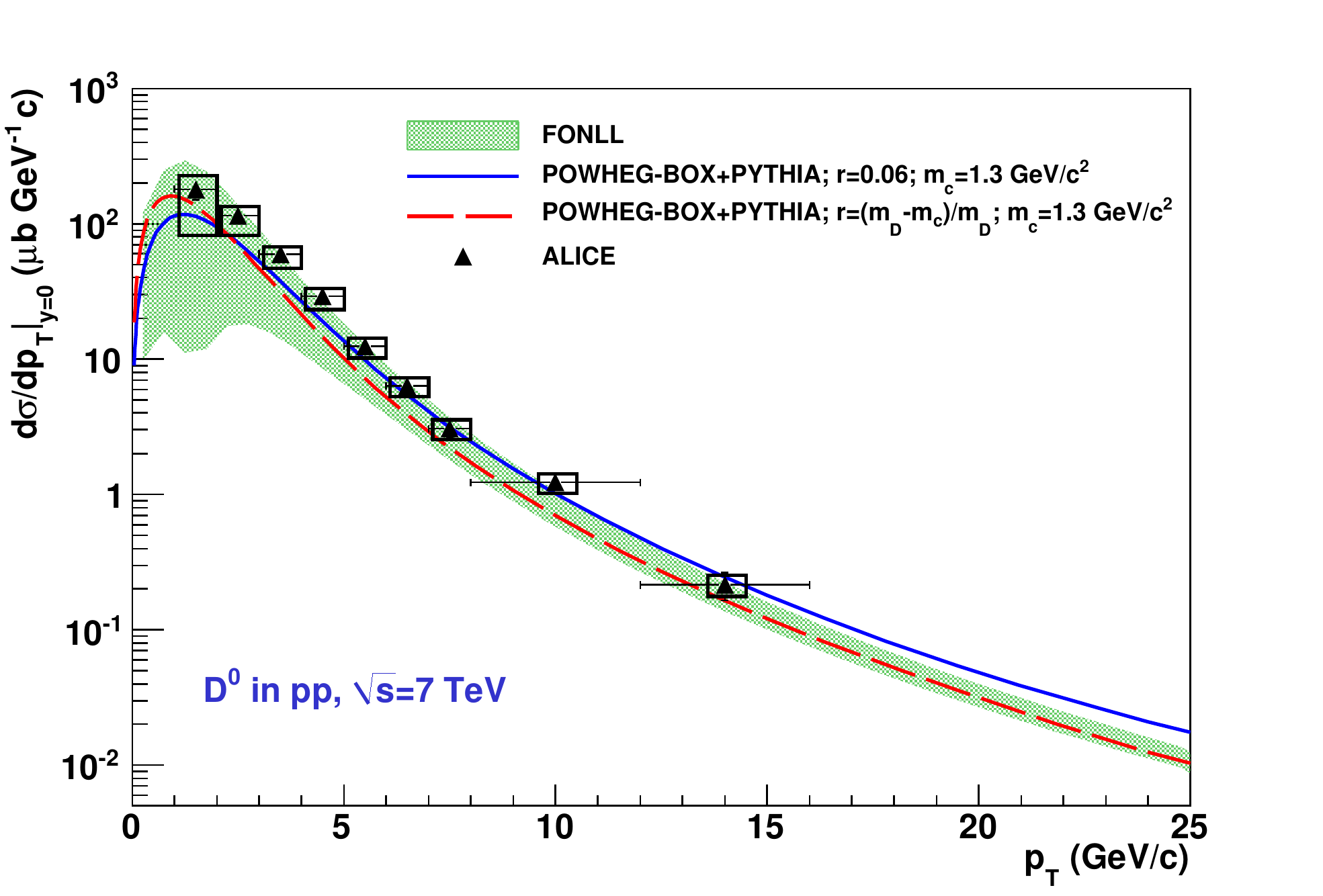}\label{fig:pp}                     
\end{center}
\caption{A cartoon of heavy-quark production within the POWHEG-BOX setup~\cite{Alioli:2010xd}, with the hard process interfaced with a parton-shower stage. Results for the final D-meson spectra in pp collisions at $\sqrt{s}\!=\!7$ TeV are displayed in the right panel~\cite{Alberico:2013bza}, compared to FONLL predictions~\cite{Cacciari:1998it} and ALICE data~\cite{ALICE:2011aa}.}
\end{figure}
Because of their large mass, heavy ($c$ and $b$) quarks are produced in hard processes occurring on a very short time-scale. Hence, in A-A collisions, their initial production is not expected to suffer any modification due to the presence of a medium, which does not have time to form and in case would not be able to affect physics at very short distances. The only initial-state effect is thought to arise from the nuclear modification of the Parton Distribution Functions (PDFs) and from the larger transverse momentum broadening of the incoming partons, which have to cross a slice of nuclear matter before colliding. Hence, the initial hard $Q\overline{Q}$ production can be thought to occur in independent nucleon-nucleon collisions and, in order to quantify later medium effects present in A-A collisions, one has to check that the pp case is under control.
The state of the art in the description of the initial $Q\overline{Q}$ production is represented by NLO pQCD calculations (POWHEG, MC@NLO) for the hard process interfaced to some event generator (PYTHIA, HERWIG) to simulate the Initial and Final State Radiation and other non perturbative processes (intrinsic $k_T$, Underlying Event and hadronization). Automated tools, such as the POWHEG-BOX package~\cite{Alioli:2010xd}, are available to perform such calculations producing fully-exclusive information of the final state. As displayed in the right panel of Fig.~\ref{fig:pp}, a satisfactory agreement with the experimental data can be obtained, both for the inclusive single-particle spectra and (not shown) for more differential observables sensitive to the initial $Q\overline{Q}$ correlations. Unfortunately, current measurements are rather indirect (e-h or D-h) and, in the A-A case, medium effects are difficult to model; actually $Q\overline{Q}$ production can occur in processes (gluon-splitting and flavor excitation) characterized by the presence of a third hard parton on the away-side, which will affect the final observables and whose propagation in the medium should be also described.

\section{Heavy flavor in nucleus-nucleus collisions}\label{sec:AA}
\begin{figure}[ht]
\begin{center}
\includegraphics[clip,width=0.51\textwidth]{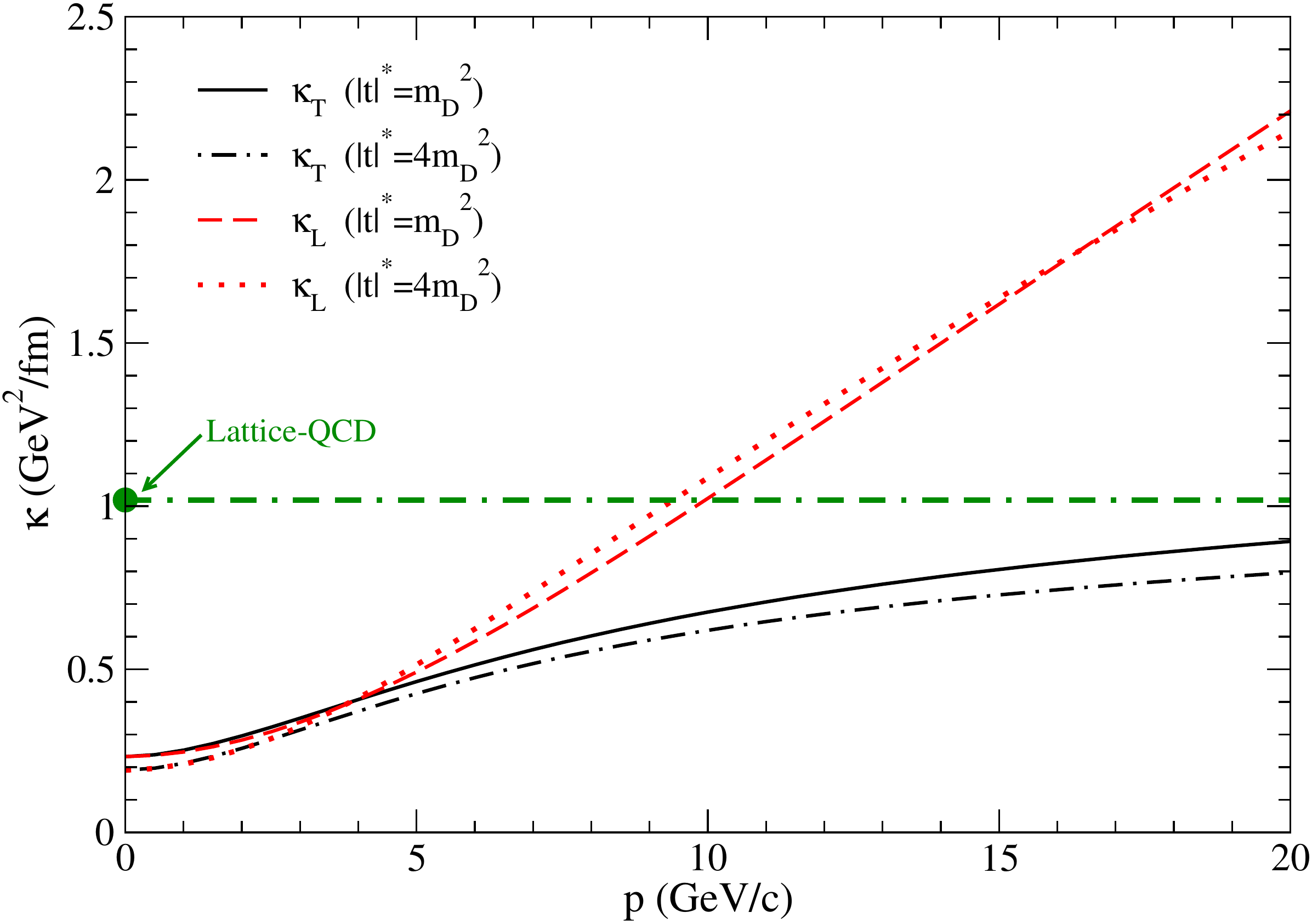}
\includegraphics[clip,width=0.45\textwidth]{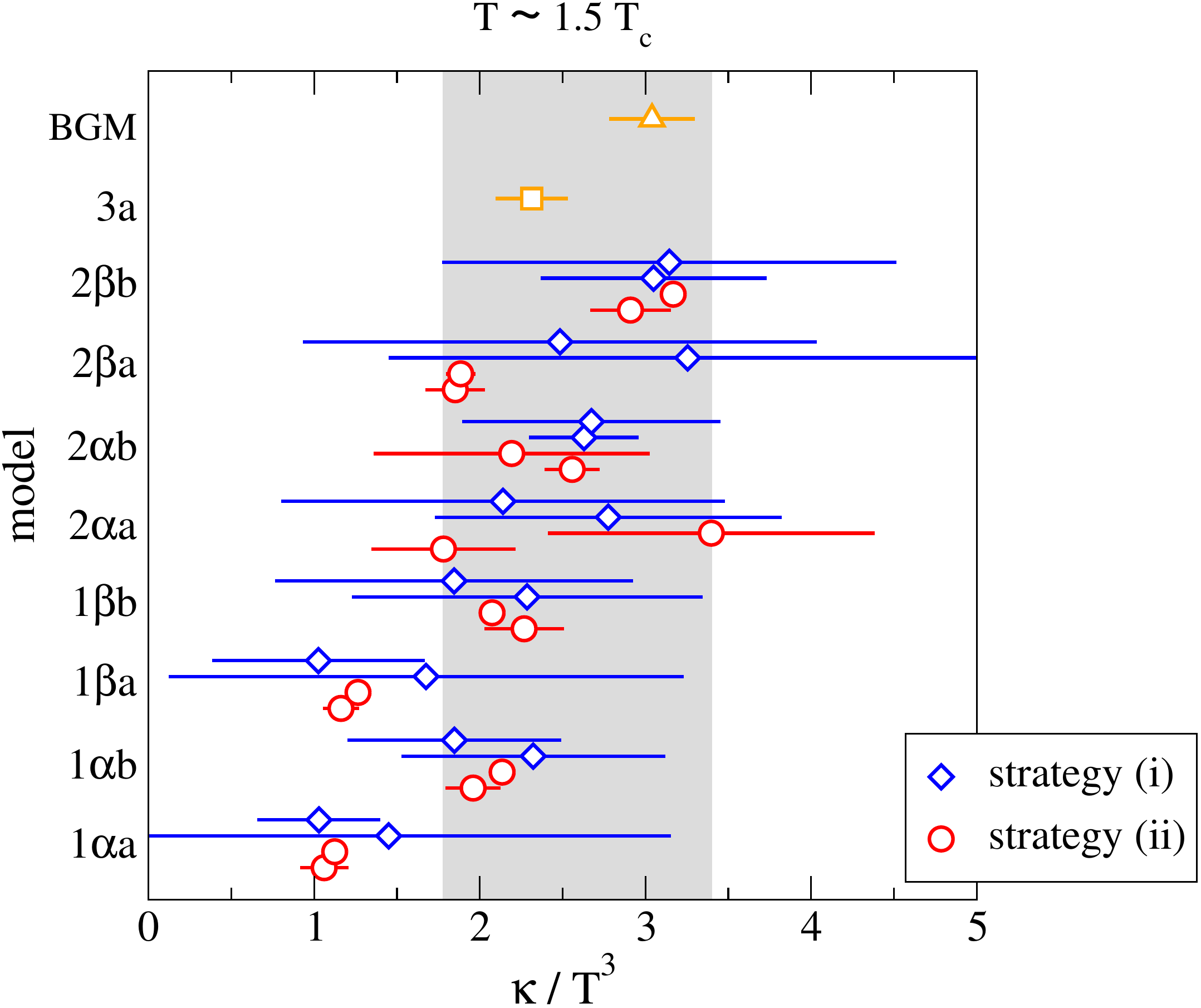}                     
\end{center}
\caption{Heavy-flavor transport coefficients. Weak coupling (HTL+pQCD) results for $b$ quarks (left panel,~\cite{Alberico:2013bza}) and the systematic uncertainty band of lattice-QCD estimates (right panel,~\cite{Francis:2015daa}) are displayed.}\label{fig:transport}
\end{figure}

Heavy quarks, whose initial hard production is quite well under control, can be used as calibrated probes of the medium formed in heavy-ion collisions, where, on average, they have to propagate for several fm/c's before hadronizing and finally decoupling. Experimental heavy-flavor measurements (D mesons, $J/\psi$'s from B decays, heavy-flavor electrons...) show that the $p_T$-spectra of charm and beauty are suppressed with respect to pp collisions (indicating an important in-medium energy loss) and provide evidence of a non-vanishing elliptic flow $v_2$ in non-central collisions: charm seems to follow the collective flow of the medium. This last observation raises the question to what extent charm quarks approach kinetic equilibrium with the rest of the medium during the limited lifetime of the latter, which furthermore undergoes a hydrodynamic expansion. Transport calculations represent the tool to study heavy-quark propagation in the QGP.  As long as the medium admits a particle description the starting point of all transport calculation is the Boltzmann equation for the evolution of the heavy quark phase-space distribution: 
\beq
{\frac{d}{dt}f_Q(t,\x,\p)=C[f_Q]}\quad{\rm with}\quad C[f_Q]=\int d\k[{w(\p+\k,\k)f_Q(t,\x,\p+\k)}-{w(\p,\k)f_Q(t,\x,\p)}]
,\label{eq:Boltzmann}
\eeq
where the collision integral $C[f_Q]$ is expressed in terms of the $\p\to\p-\k$ transition rate $w(\p,\k)$
The direct solution of the Boltzmann equation in the evolving inhomogeneous fireball produced in heavy-ion collisions is challenging; results can be found in~\cite{Gossiaux:2008jv,Uphoff:2011ad}. However as long as $k\ll p$ ($k$ being typically of order $gT$) one can expand the collision integral in powers of the momentum exchange.
 Relying on the above mentioned hypothesis of dominance of soft scatterings one can approximate the exact Boltzmann equation with a form more suited to an easy numerical implementation: the relativistic Langevin equation. The latter allows one to study the evolution of the momentum $\vec{p}$ of each heavy quark in the medium, through the combined effect of a friction and a noise term, both arising from the collisions suffered in the plasma:
\begin{equation}
  \frac{\Delta\vec{p}}{\Delta t}=-\eta_D(p)\vec{p}+\vec{\xi}(t)\;\;{\rm with}\;\; \langle\xi^i(t)\xi^j(t')\rangle\!=\!b^{ij}(\vec{p})\delta_{tt'}/\Delta t\!\equiv\! \left[\kappa_L(p)\hat{p}^i\hat{p}^j\!+\kappa_T(p)\!
(\delta^{ij}-\hat{p}^i\hat{p}^j)\right]\delta_{tt'}/\Delta t.
\label{eq:lange_r_d}
\end{equation}
 The strength of the noise is set by the momentum-diffusion coefficients $\kappa_{L/T}$, reflecting the average longitudinal/transverse squared momentum exchanged with the plasma. The friction coefficient $\eta_D(p)$ must be fixed in order to ensure the approach to thermal equilibrium by the Einstein fluctuation-dissipation relation
\begin{equation}
\eta_D(p)\equiv\frac{\kappa_L(p)}{2TE_p}+{\rm corrections},\label{eq:friction}
\end{equation}
where the corrections, subleading by a factor ${\cal O}(T/E_p)$, depend on the discretization scheme and are fixed in order to reproduce the correct continuum result in the $\Delta t\to 0$ limit. The advantage of the Langevin approach is that, independently of the nature of the medium (even when the latter does not admit a quasi-particle description), it allows one to summarize the interaction of the heavy quarks with the plasma in a few transport coefficients with a clean physical interpretation. Results for the momentum-diffusion coefficients, obtained with very different approaches (resummed weak-coupling calculations and lattice-QCD simulations) are displayed in Fig.~\ref{fig:transport}. One finds that, at zero momentum, the lattice result is much larger than the perturbative value; however weak-coupling calculations provide evidence, in particular for the case (not shown) of charm, of a quite strong momentum dependence of $\kappa_{L/T}$ (the latter is found to be even larger in other approaches, like AdS/CFT calculations~\cite{Gubser:2006nz}), which unfortunately at present cannot be accessed on the lattice. This represents a limit on the information one can extract from experimental charm measurements if one cannot access the low-$p_T$ region. On the other hand, one can see how, for beauty, $\kappa_{L/T}$ values stay close to each other and quite constant for an extended momentum range, opening the possibility with the ALICE upgrade to perform measurements able to tightly constrain the value of the QGP transport coefficients. 
\begin{figure}[ht]
\begin{center}
\includegraphics[clip,height=5.5cm]{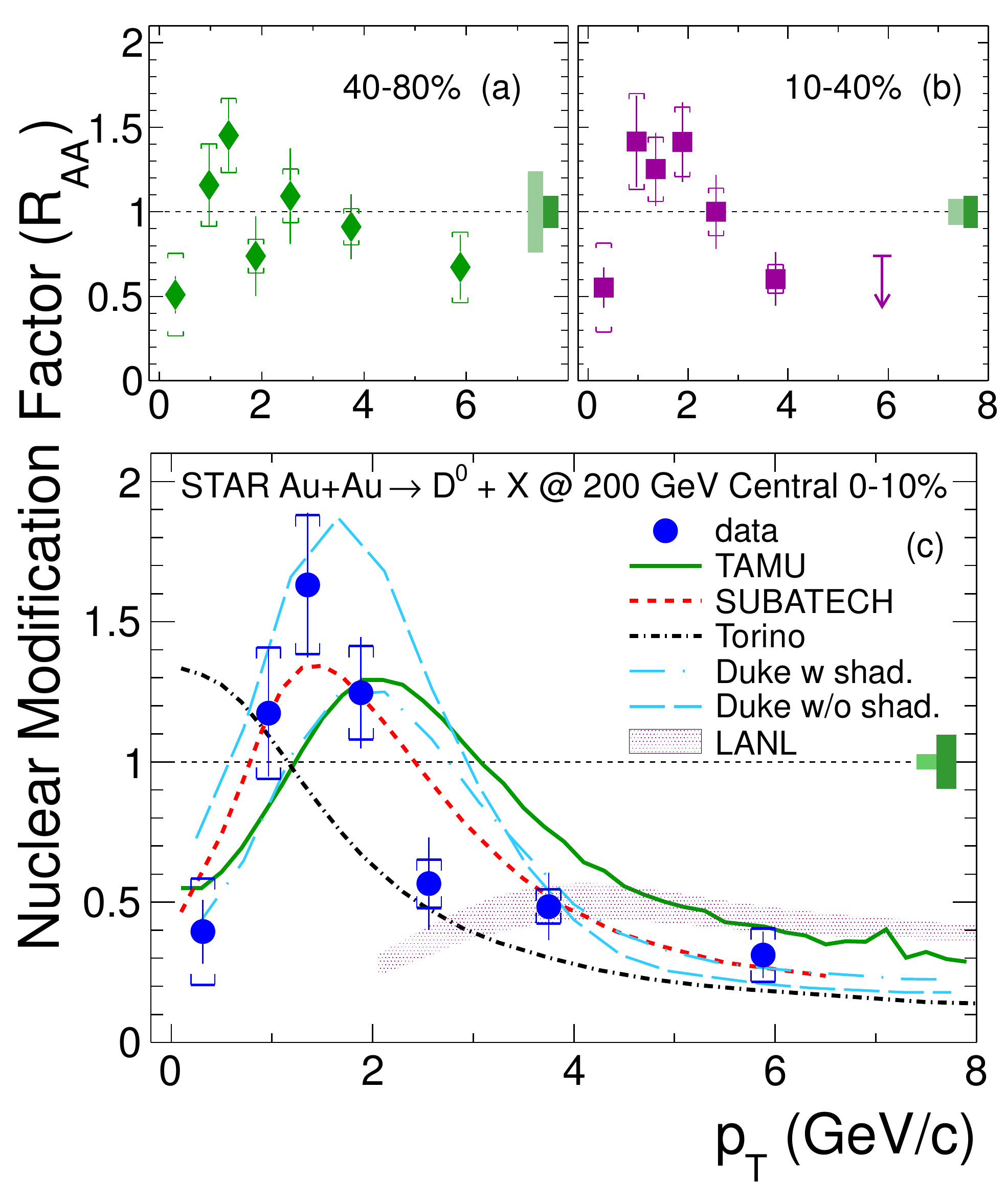}
\includegraphics[clip,height=5.5cm]{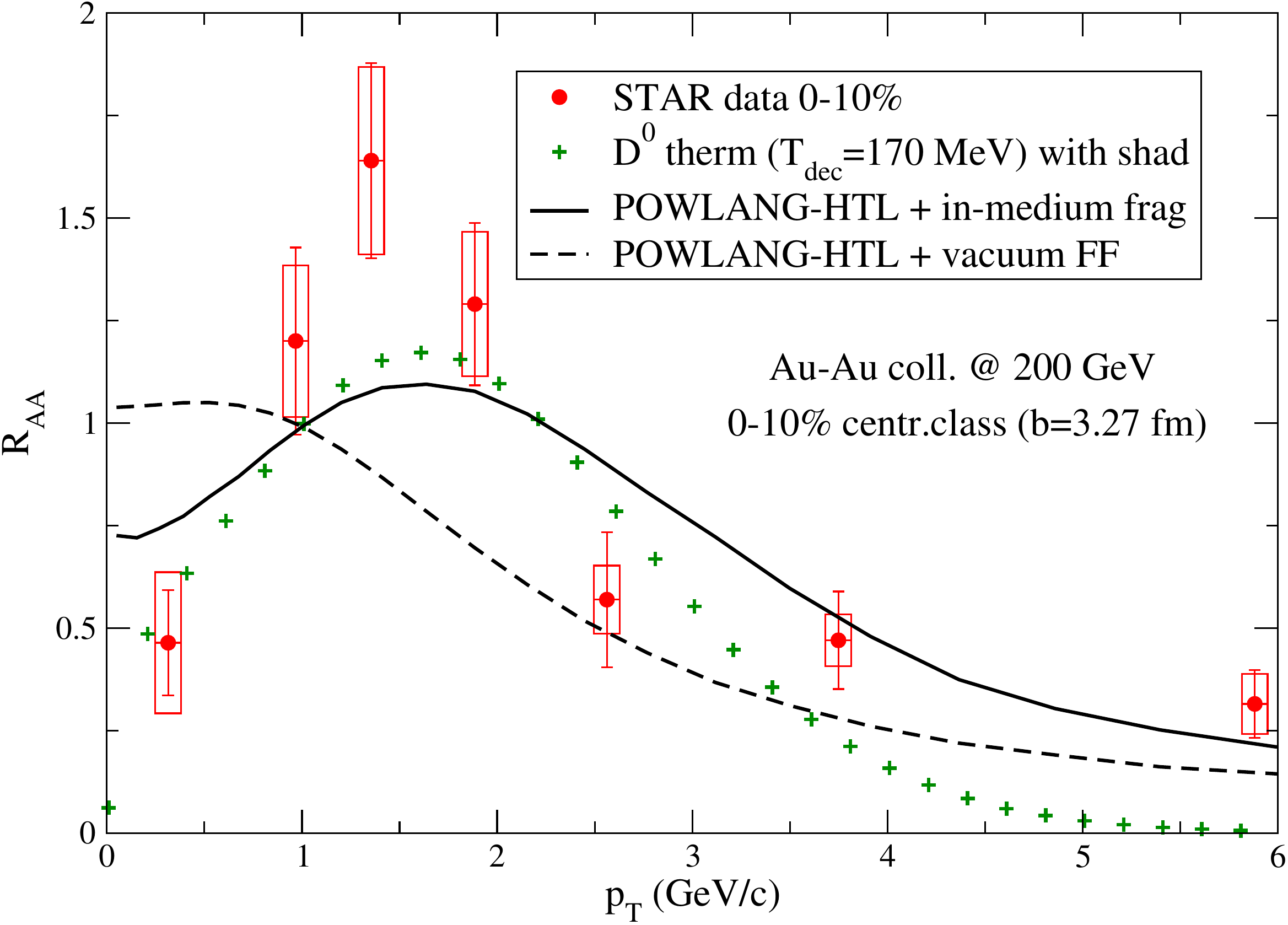}
\end{center}
\caption{The D-meson $R_{\rm AA}$ in central Au-Au collisions at $\sqrt{s_{\rm NN}}\!=\!200$ GeV. Left panel: STAR results~\cite{Adamczyk:2014uip} are compared to various model calculations. Right panel: POWLANG results (black curves) with and without in-medium hadronization, together with the limit of full kinetic equilibrium (green crosses), are displayed~\cite{Beraudo:2014boa}.}\label{fig:RAA}
\end{figure}

The presence of a deconfined medium, besides modifying the heavy quark propagation, can also affect their hadronization, changing both the relative particle abundances and their momentum spectra. While in the vacuum hadrons come from the fragmentation of partons produced in the hard event, in A-A collisions recombination with light thermal partons may be at work. Since, in particular, the environment is very rich in strange quarks, changes in the heavy-flavor hadrochemistry may be expected, leading for instance to an enhanced production of $D_s$ mesons~\cite{Innocenti:2012ds}. The effect of heavy-quark coalescence with light partons on the final particle spectra was studied in detail in~\cite{vanHees:2007me}, where it turned out to allow a better description of the $R_{\rm AA}$ and $v_2$ of heavy-flavor electrons at RHIC. In the left panel of Fig.~\ref{fig:RAA} it is shown how models including coalescence display a better agreement with STAR data~\cite{Adamczyk:2014uip} for the D-meson $R_{\rm AA}$.

\begin{figure}[ht]
\begin{center}
\includegraphics[clip,width=0.48\textwidth]{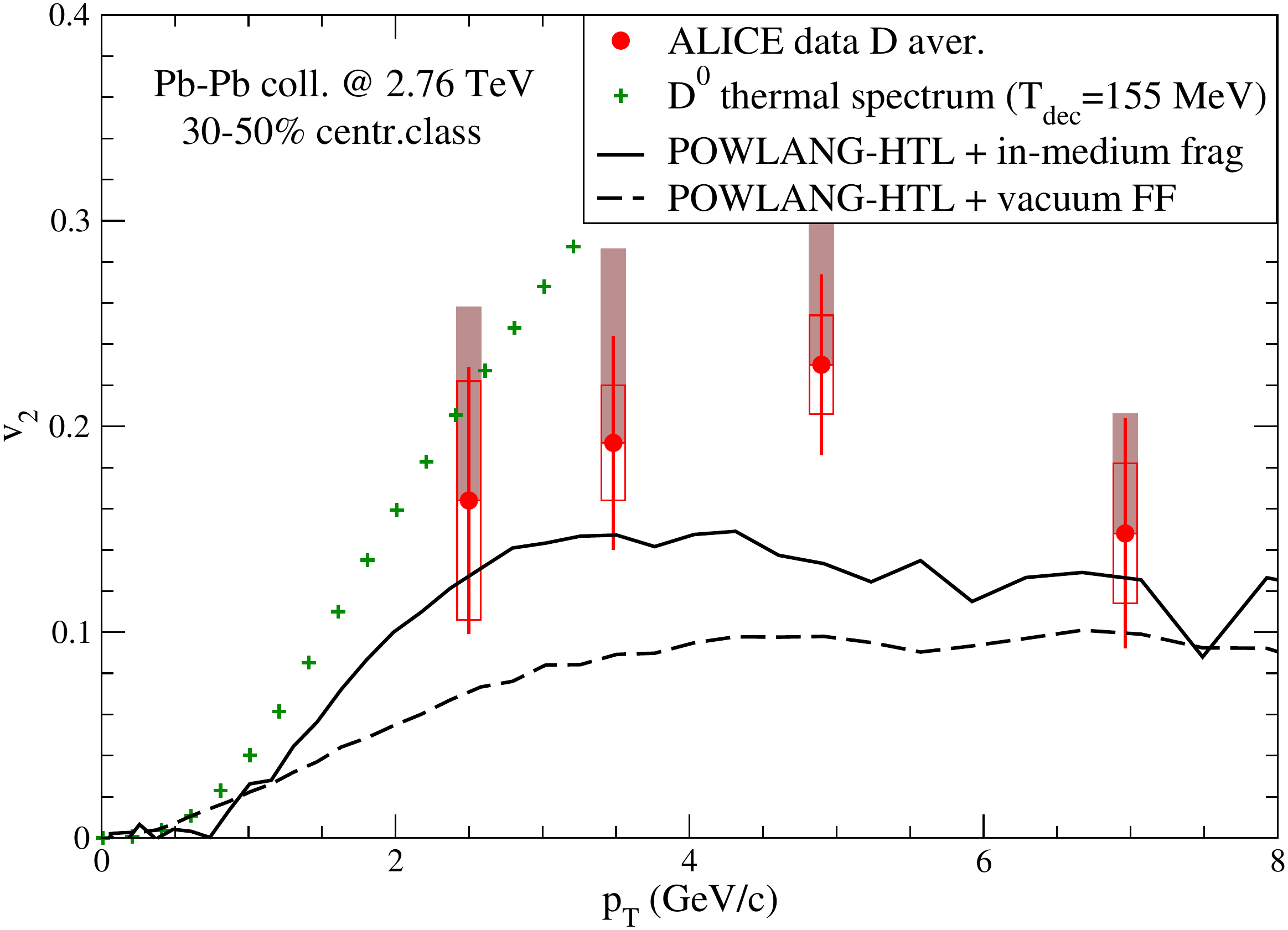}
\includegraphics[clip,height=5.5cm]{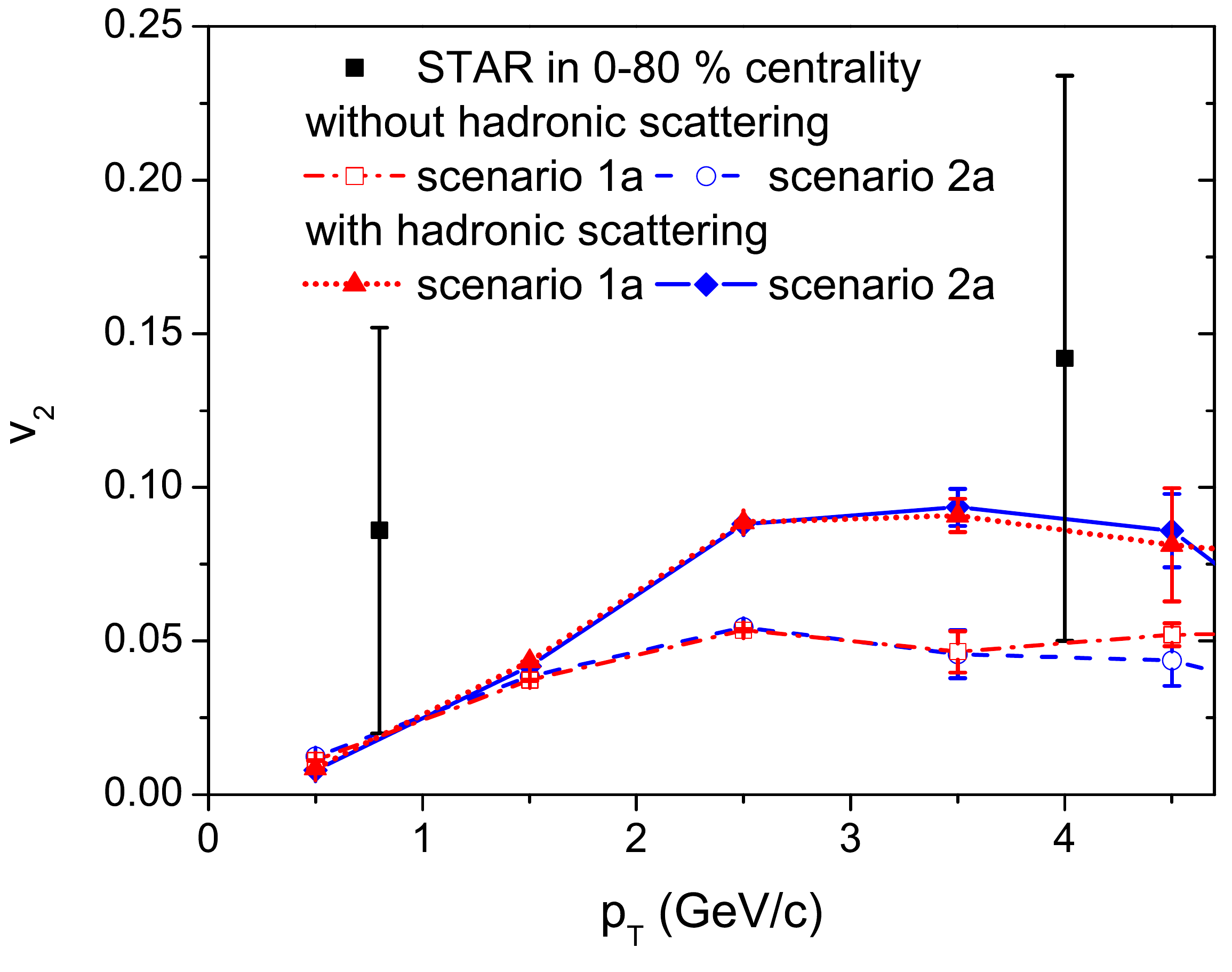}
\end{center}
\caption{D-meson elliptic flow. Left panel: POWLANG results (black curves) with and without in-medium hadronization, together with the limit of full kinetic equilibrium (green crosses), are shown~\cite{Beraudo:2014boa} and compared to ALICE data in non-central Pb-Pb collisions at the LHC~\cite{Abelev:2013lca}. Right panel: the relevance of the possible D-meson rescattering in the hadronic phase is displayed~\cite{Song:2015sfa}. Results are compared to STAR data in minimum-bias Au-Au collisions~\cite{Tlusty:2012ix}.}\label{fig:v2}
\end{figure}
In the coalescence picture, the recombination probability is large when the wave-function of the final D meson (if one considers charm) and the wave-packets of the two partons (the $c$ quark and a light antiquark from the medium) display a sizable overlap; this occurs when the initial partons are sufficiently close in space and have comparable velocities.
An alternative picture is the one recently implemented in the POWLANG setup~\cite{Beraudo:2014boa}. Here, heavy quarks, once they reach a cell below a decoupling temperature, independently of their kinematics, are combined with light thermal partons from the medium, giving rise to $Q\overline{q}$ strings eventually fragmented according to the PYTHIA~\cite{Sjostrand:2006za} implementation of the Lund model into the final charmed (or beauty) hadrons. Although the mechanism is slightly different from coalescence, the result is similar: heavy-flavor hadrons tend to inherit part of the collective flow of the medium from the light thermal partons, leading to a better agreement with the data.
This can be appreciated in the right panel of Fig.~\ref{fig:RAA}, where STAR data for the D-meson $R_{\rm AA}$ are compared to different POWLANG predictions. As it can be seen, with the same transport coefficients, results including in-medium hadronization better reproduce the qualitative trend of the data, with the peak at moderate $p_T$ arising from radial flow. The same occurs for the elliptic flow, as displayed in the left panel of Fig.~\ref{fig:v2} where ALICE data~\cite{Abelev:2013lca} in non-central Pb-Pb collisions are plotted together with POWLANG predictions: also in this case the additional flow acquired via recombination with light partons from the medium moves theory outcomes closer to the experimental findings.

Finally, although very often neglected, heavy-flavor hadrons can go on scattering also in the hadronic phase. Effective chiral lagrangians can be employed to describe the interactions between D mesons and the other light particles like pions or kaons~\cite{Song:2015sfa,e:2011yi}. Of course the hadronic phase is characterized by a lower value of the the temperature and hence of the transport coefficients compared to the QGP, but it is also the stage in which the system has acquired the largest flow: a non-negligible $v_2$ can be acquired also in the hadron gas, as show in the right panel of Fig.~\ref{fig:v2}. 
\section{Heavy flavor in small systems}\label{sec:pA}
\begin{figure}[ht]
\begin{center}
\includegraphics[clip,height=5.5cm]{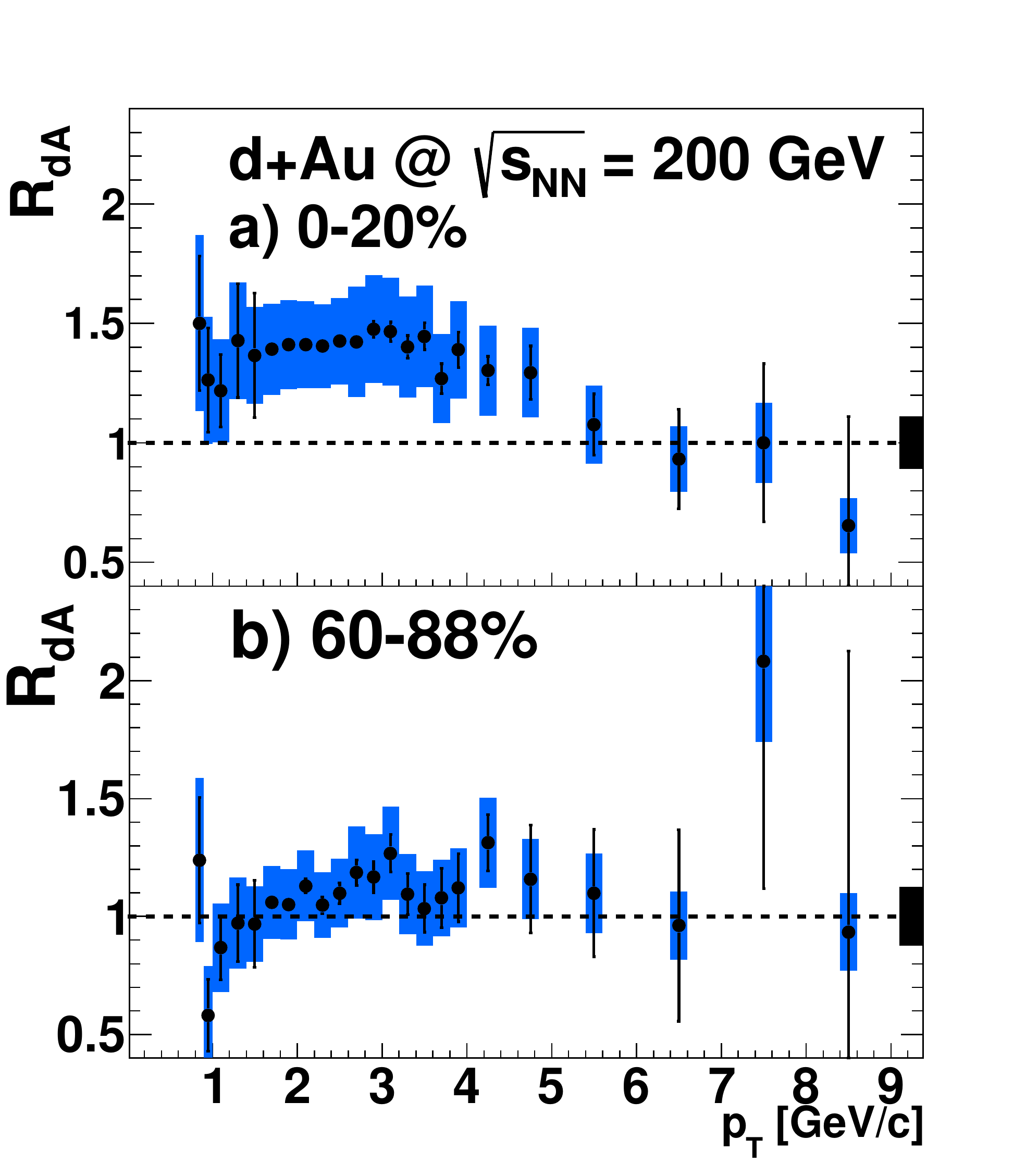}
\includegraphics[clip,height=5cm]{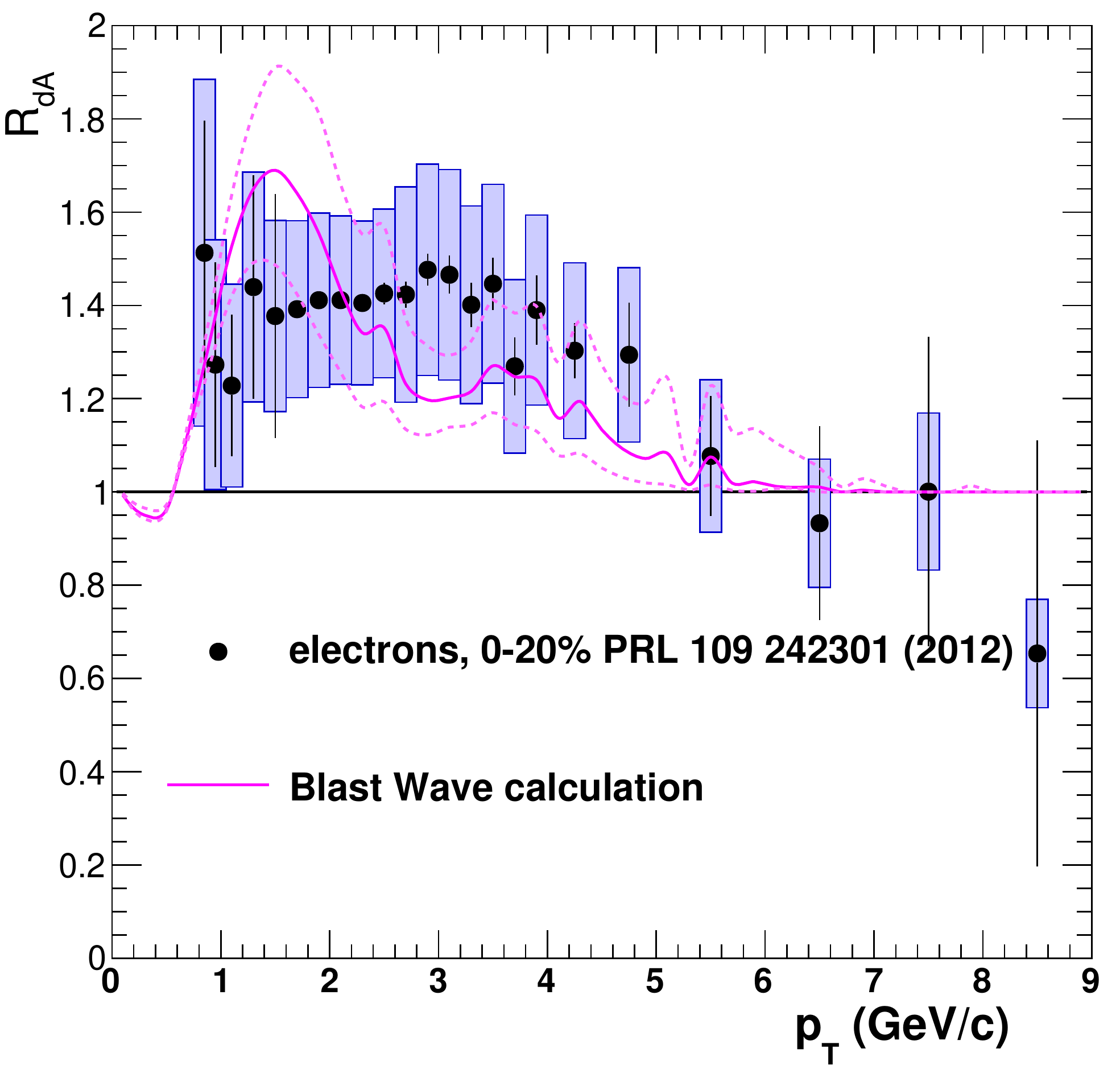}
\includegraphics[clip,height=5.2cm]{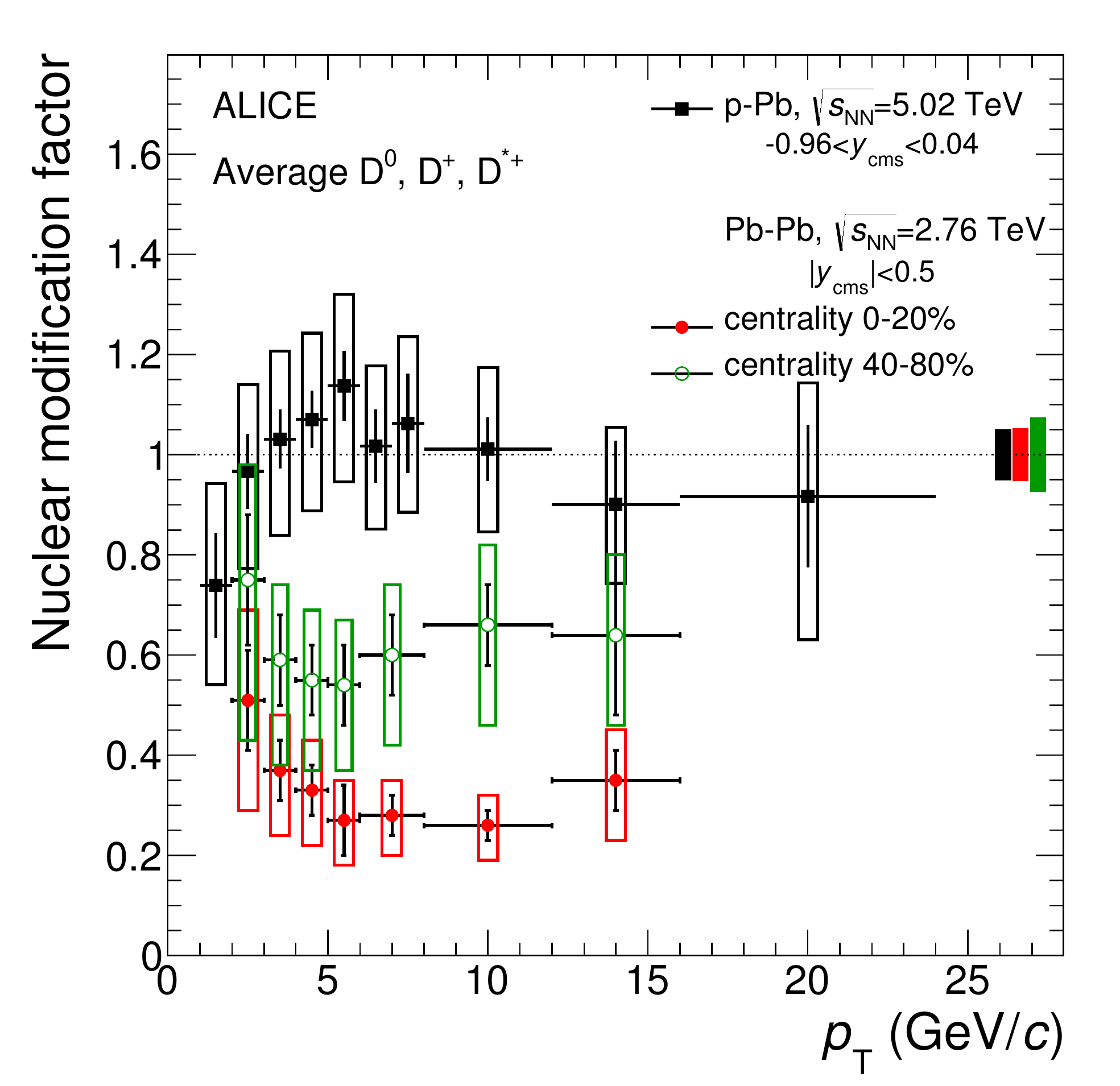}                     
\end{center}
\caption{Heavy-flavor nuclear modification in small systems. Left and middle panels: PHENIX results for heavy-flavor electrons in central d-Au collisions at RHIC~\cite{Adare:2012yxa} compared to peripheral events and to a blast-wave fit~\cite{Sickles:2013yna}, assuming a common collective flow with the medium. Right panel: ALICE results for D-mesons in minimum-bias p-Pb collisions at the LHC~\cite {Abelev:2014hha}, compared to central and peripheral Pb-Pb events~\cite{ALICE:2012ab}.}\label{fig:RpA}
\end{figure}
\begin{figure}[ht]
\begin{center}
\includegraphics[clip,width=0.48\textwidth]{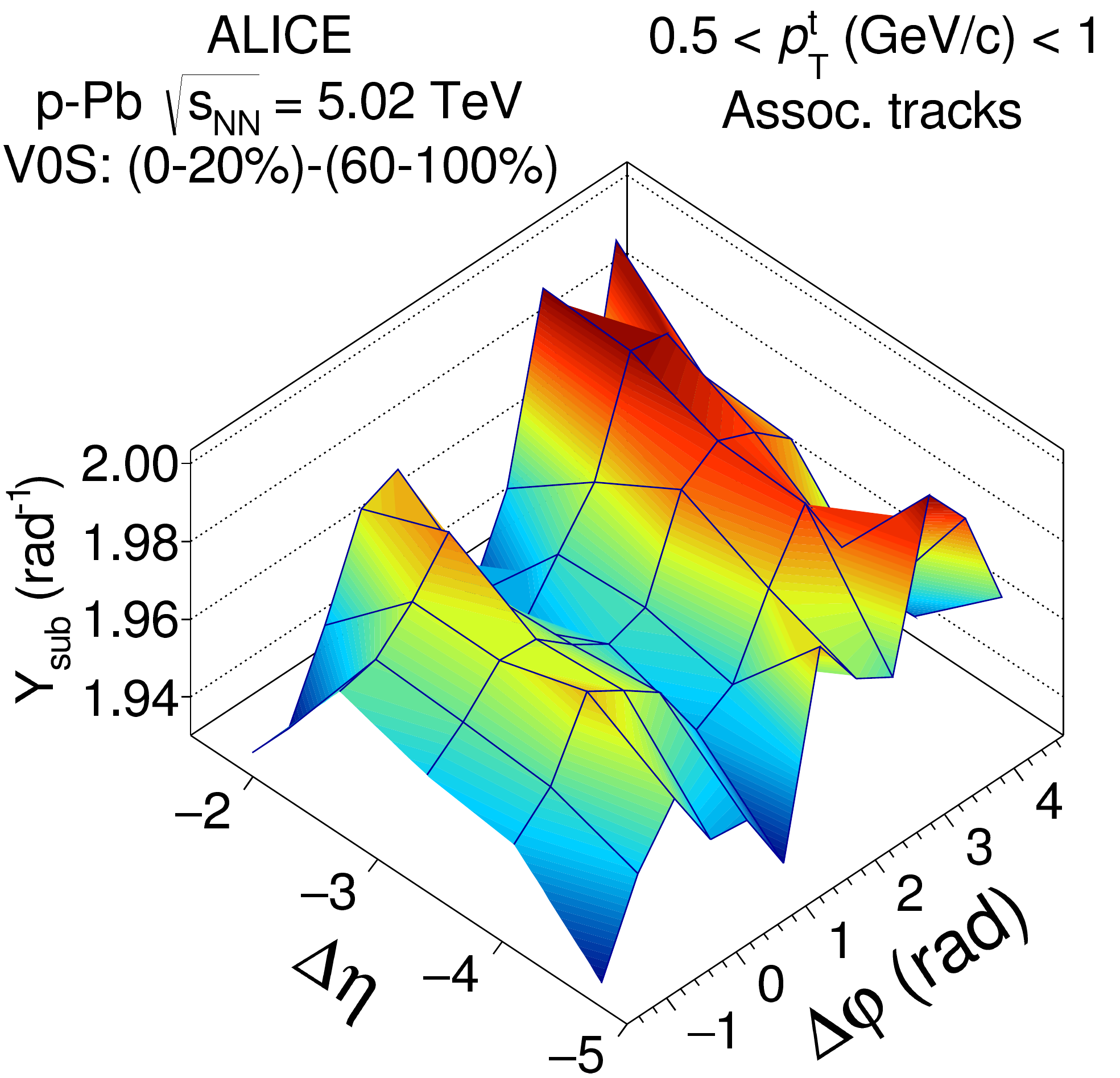}
\includegraphics[clip,width=0.48\textwidth]{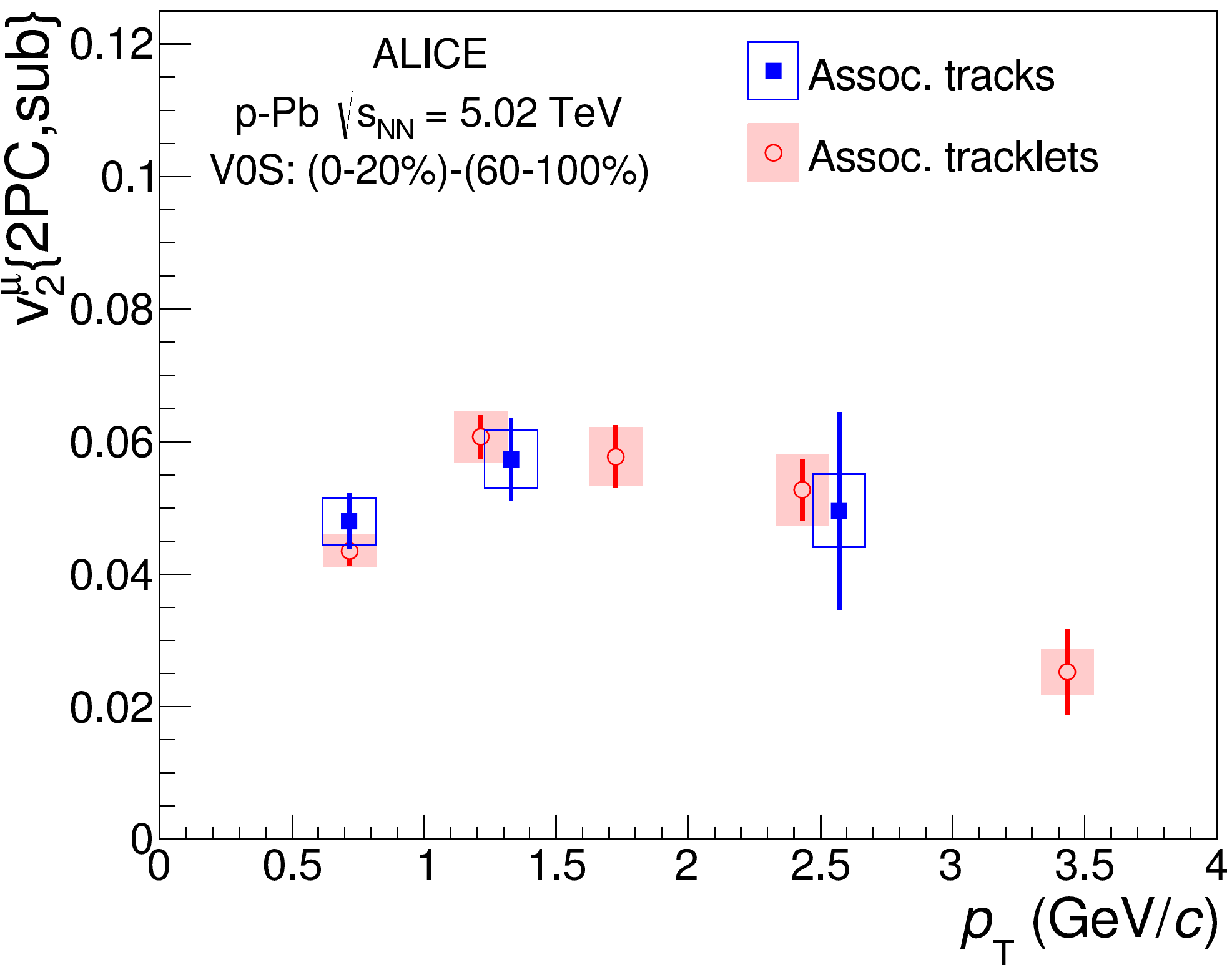}                     
\end{center}
\caption{ALICE results for muon-hadron correlations in p-Pb collisions, after subtraction of the results in peripheral events~\cite{Adam:2015bka}. The corresponding estimate of the muon elliptic flow is displayed in the left panel.}\label{fig:mu-h}
\end{figure}
One of the surprises in the experimental search for QGP formation came from the observation of collective effects in small systems: the ridge in long-range rapidity correlation in high-multiplicity pp and p-Pb events, the non-vanishing flow harmonics measured in p-Pb and d-Au collisions, the $p/\pi$ ratio of the $p_T$-spectra, etc. All the above effects involve soft light hadrons and suggest the formation of a hot medium with a hydrodynamic behavior even in these collisions. On the other hand, no significant effect was observed in the high-$p_T$ domain; neither jets nor particle spectra turn out to be suppressed and minor modifications can be interpreted as an initial-state effect due to the nuclear PDFs. Actually, the two observations are not necessarily in contradiction. In spite of the possible strongly interacting nature of the medium (responsible for the hydrodynamic expansion driven by pressure gradients), the energy loss of hard partons depends in any case on the system size ($\langle\Delta E\rangle\!\sim\!L^2$ in the presence of coherence effects).

One may wonder what happens to heavy-flavor production in these small systems. The situation is not completely settled.
At RHIC (see Fig.~\ref{fig:RpA}) the nuclear modification factor of heavy-flavor electrons in central d-Au collisions was found to slightly exceed 1~\cite{Adare:2012yxa}. Although error bars are large this looks puzzling; a possible explanation was proposed in Ref.~\cite{Sickles:2013yna}, where a blast-wave spectrum for D and B mesons was employed, assuming they share the same collective flow of the medium.
At the LHC the D-meson $R_{\rm pA}$ in minimum-bias p-Pb collisions turns out to be compatible with 1 within the large systematic error-bars~\cite{Abelev:2014hha}, with a depletion at low $p_T$ to attribute to the nPDFs. On the other hand, other observations suggest signatures of collective effects also in the heavy-flavor sector, such as (see Fig.~\ref{fig:mu-h}) the appearance of a double ridge in $e$-h and $\mu$-h correlations~\cite{Adam:2015bka} measured in central p-Pb events, where electrons and (part of the) muons come from heavy-flavor decays, and from which it is possible to extract an estimate of the elliptic-flow of the trigger particle. 

\begin{figure}[ht]
\begin{center}
\includegraphics[clip,height=5.4cm]{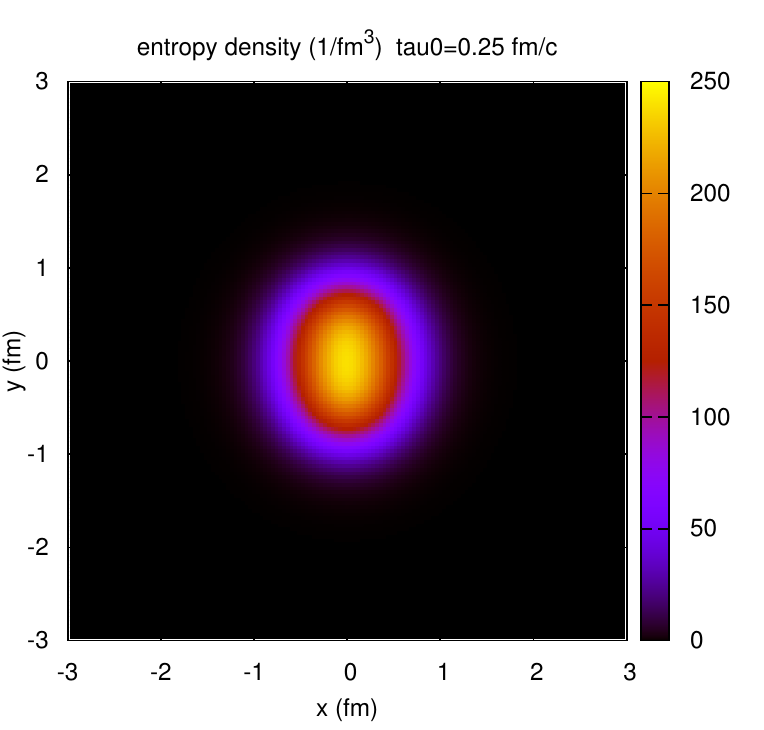}
\includegraphics[clip,height=5cm]{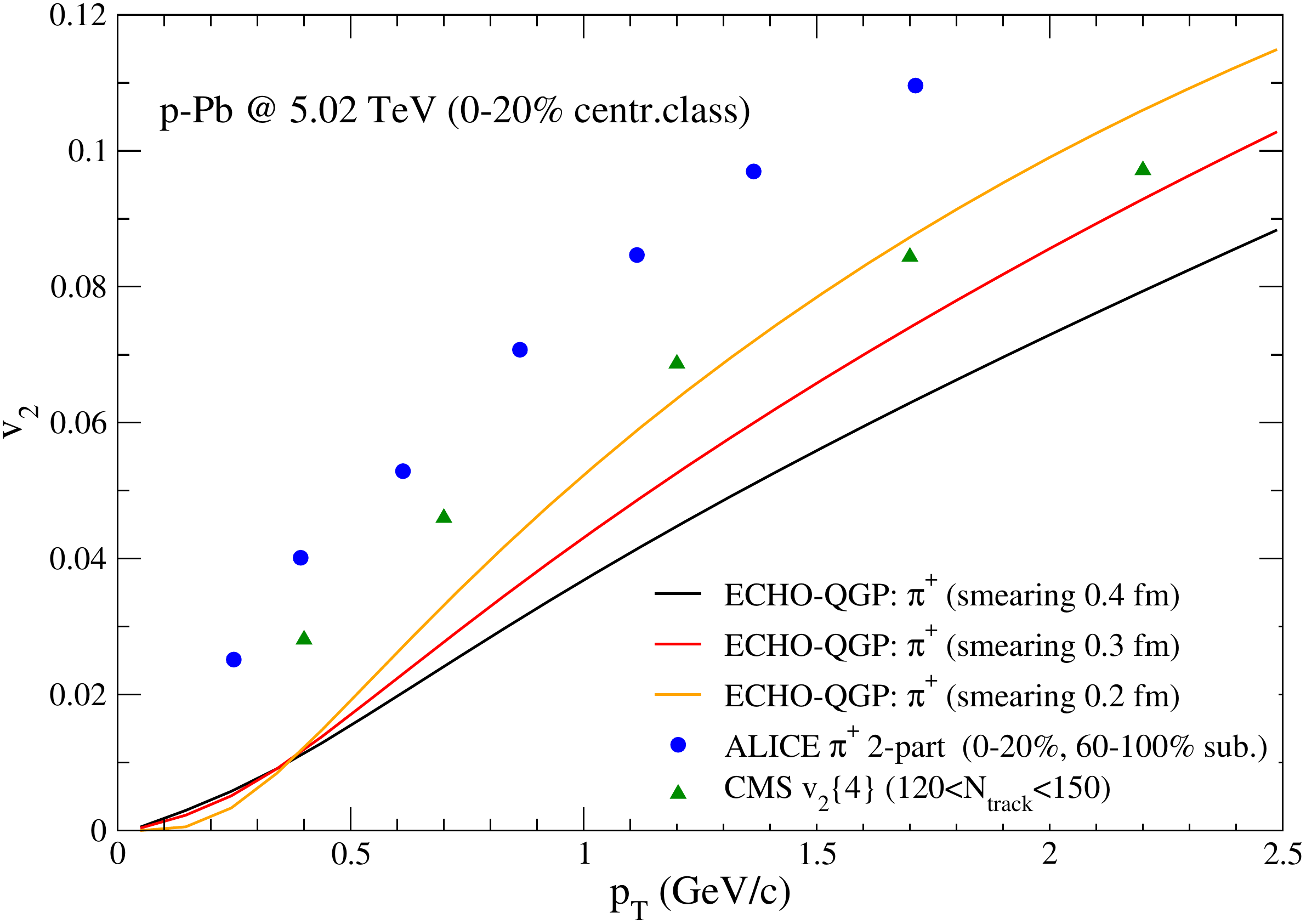}                     
\end{center}
\caption{Average initial condition for 0-20\% more central p-Pb events at $\sqrt{s_{\rm NN}}\!=\!5.02$ TeV and the corresponding pion elliptic flow (for various smearing parameters) arising from the hydrodynamic evolution of the medium compared to ALICE and CMS data.}\label{fig:hydro}
\end{figure}
\begin{figure}[!t]
\begin{center}
\includegraphics[clip,height=5.5cm]{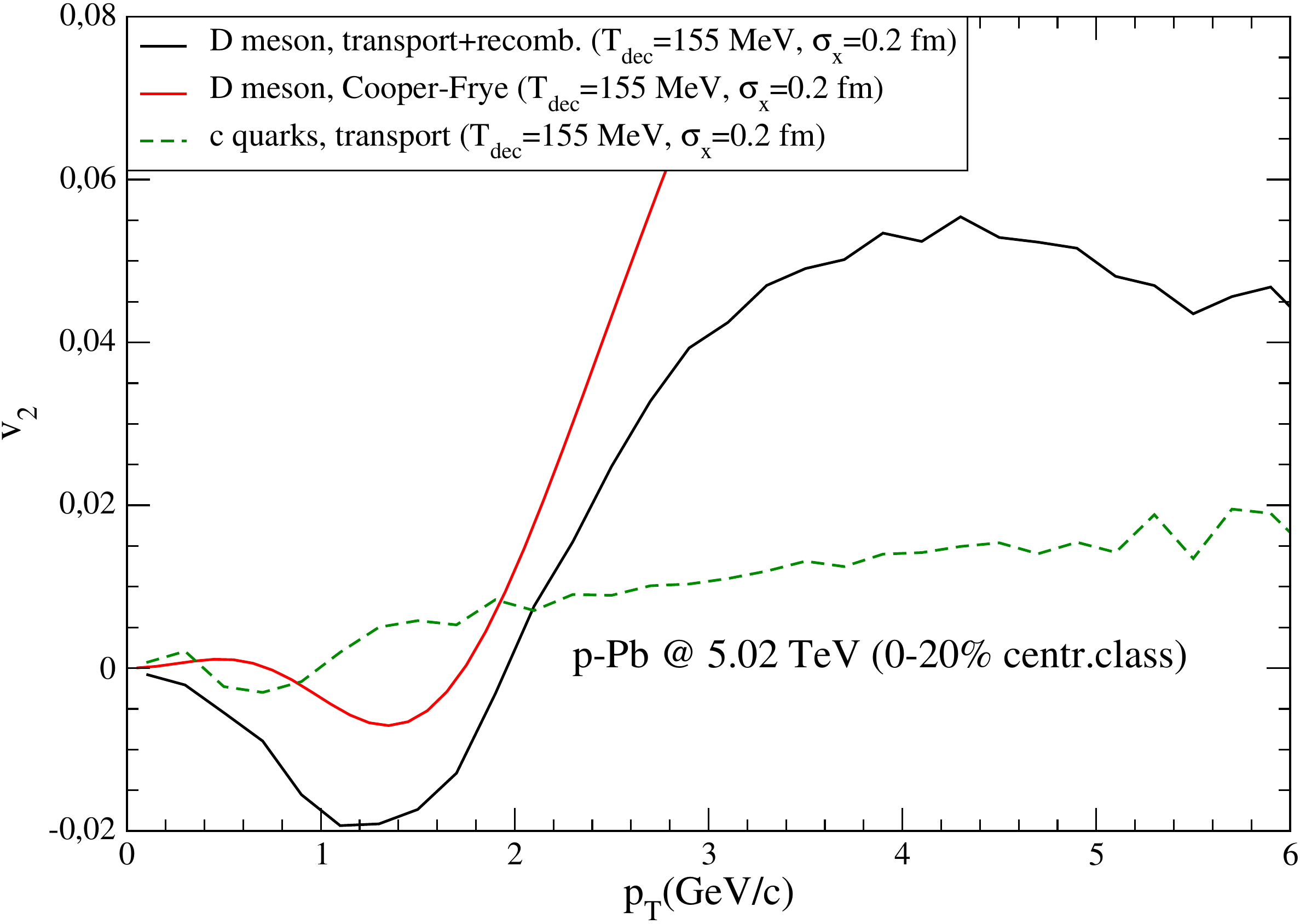}
\includegraphics[clip,height=6cm]{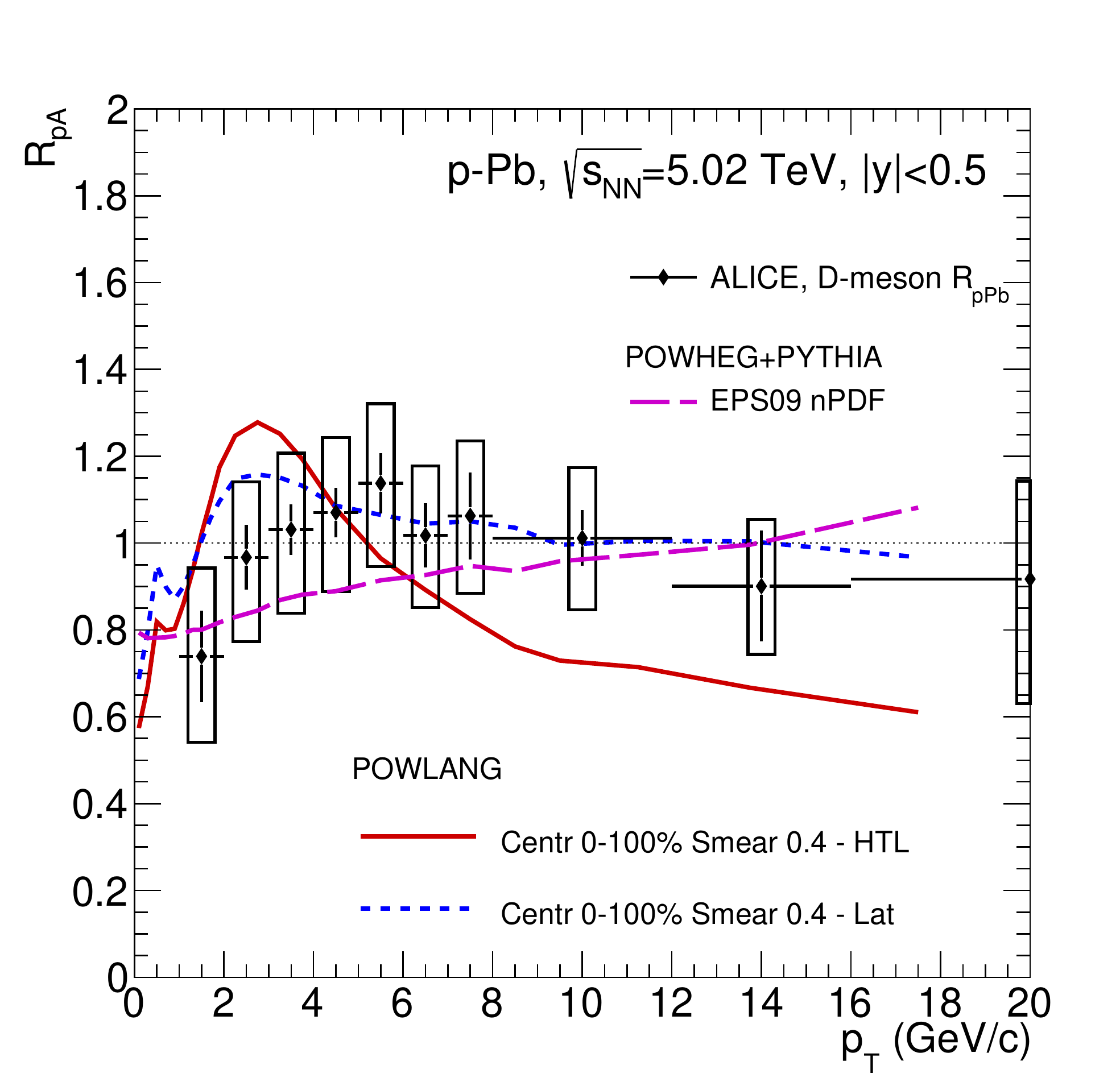}
\end{center}
\setlength{\abovecaptionskip}{-4pt}
\setlength{\belowcaptionskip}{-4pt}
\caption{Left panel: POWLANG results for charm elliptic flow in 0-20\% more central p-Pb events. The $v_2$ of $c$ quarks, D-meson (after in-medium hadronization) as well as the limit of full kinetic equilibrium are compared. Right panel: POWLANG predictions for the D-meson nuclear modification factor in minimum-bias p-Pb collisions.}\label{fig:pAresults}
\end{figure}

Can the propagation and subsequent hadronization of heavy quarks in the presence of a small-size hot deconfined medium provide results compatible with the present data? We attempt to model the situation as follows (see Ref.~\cite{Nardi:2015pca} for details). A smooth hydrodynamic background is constructed summing (depending on the centrality selection) hundreds/thousands Glauber-MC fluctuating initial conditions, each one rotated by the event-plane angle $\psi_2$ so as to get a realistic estimate of the average initial eccentricity $\epsilon_2$; the resulting profile 
is then used as an initial condition for the numerical solution of viscous hydrodynamic equations performed with the ECHO-QGP code~\cite{DelZanna:2013eua}. As shown in Fig.~\ref{fig:hydro}, the procedure allows one to roughly reproduce the size of the light-hadron elliptic flow, so that one can be confident to have a quite realistic description of the medium. The solution of hydrodynamic equations can then be used as a background for the heavy quark propagation, simulated through the POWLANG setup. As done in the A-A case, below a certain decoupling temperature $c$ quarks are then combined with thermal partons to form strings whose fragmentation will give rise to the final D-mesons. Results are shown in Fig.~\ref{fig:pAresults}. As the crossed medium is very thin, the elliptic flow of $c$ quarks at decoupling is very small; however, recombination with light thermal partons may enhance the $v_2$ up to $\sim$5\% because of the additional flow inherited from the medium, producing a result in qualitative agreement with the kinetic equilibrium scenario. Also for what concerns the $p_T$-spectrum, the recombination with the light background quarks leaves its fingerprints, giving rise to a bumb because of radial flow. With the present experimental error bars such an interpretation is not in contradiction with data.

\section{Conclusions and perspectives}
Heavy-flavor observables allow one to get rich information on the medium produced in heavy-ion collisions. In order to describe the data one needs to properly model both the dynamics and the hadronization of heavy quarks in the presence of a QGP. Recently, signatures of final-state medium effects were observed also in small systems (like in p-A collisions), previously considered just a benchmark to estimate cold nuclear-matter effects. This surely deserves deeper investigation in the future, together with the study of heavy-flavor correlations and of beauty at low $p_T$, allowing a more solid extraction of the heavy-flavor transport coefficients.  

\section*{References}

\end{document}